\documentclass{webofc}
\usepackage[varg]{txfonts}   % Web of Conferences font
\begin{document}
%
%===============================================================================
%
% Reference to equations, figures, tables, sections...
\newcommand{\eq}[1]   {(\ref{#1})}
\newcommand{\Eq}[1]   {Eq.~(\ref{#1})}
\newcommand{\Eqs}[2]  {Eqs.~(\ref{#1}) and~(\ref{#2})}
\newcommand{\Eqr}[2]  {Eqs.~(\ref{#1})--(\ref{#2})}
\newcommand{\Fi}[1]   {Fig.~\ref{#1}}
\newcommand{\Fis}[2]  {Figs.~\ref{#1} and~\ref{#2}}
\newcommand{\Fiss}[3] {Figs.~\ref{#1},~\ref{#2}, and~\ref{#3}}
\newcommand{\Ta}[1]   {Tab.~\ref{#1}}
\newcommand{\Tas}[2]  {Tabs.~\ref{#1} and~\ref{#2}}
\newcommand{\Tass}[3] {Tabs.~\ref{#1},~\ref{#2}, and~\ref{#3}}
\newcommand{\Se}[1]   {Sect.~\ref{#1}}
\newcommand{\Ap}[1]   {Appendix~\ref{#1}}
%
% GeV and Cie. Use only in text mode.
%\newcommand{\agev}    {\mbox{~$A$GeV}}               %Common notation
\newcommand{\agev}    {\mbox{$A$~GeV}}               %PRL notation
\newcommand{\gevc}    {\mbox{GeV$/c$}}
\newcommand{\gevcsq}  {\mbox{GeV$^{\:2}/c^{2}$}}
\newcommand{\gevcc}   {\mbox{GeV$/c^2$}}
\newcommand{\mevc}    {\mbox{MeV$/c$}}
\newcommand{\mevcc}   {\mbox{MeV$/c^2$}}
\newcommand{\mum}     {\mbox{$\mu$m}}
\newcommand{\invmub}  {\mbox{$\mu$b$^{-1}$}}
%
% Small roman for sub and superscripts
\newcommand{\rb}[1]   {\mbox{\textrm{\scriptsize #1}}}
\newcommand{\rbt}[1]  {\mbox{\textrm{\tiny #1}}}
\newcommand{\rsbox}[1]{\raisebox{1.5ex}[-1.5ex]{#1}}
\newcommand{\vzero}   {\ensuremath{\textrm{V}^{0}}}
\newcommand{\lam}     {\ensuremath{\Lambda}}
\newcommand{\lab}     {\ensuremath{\bar{\Lambda}}}  
\newcommand{\xim}     {\ensuremath{\Xi^{-}}}
\newcommand{\xizero}  {\ensuremath{\Xi^{0}}}
\newcommand{\xip}     {\ensuremath{\bar{\Xi}^{+}}}
\newcommand{\xib}     {\ensuremath{\bar{\Xi}^{0}}}
\newcommand{\sig}     {\ensuremath{\Sigma^{0}}}
\newcommand{\sip}     {\ensuremath{\Sigma^{+}}}
\newcommand{\sib}     {\ensuremath{\bar{\Sigma}^{0}}}
\newcommand{\pizero}  {\ensuremath{\pi^{0}}}
\newcommand{\pimin}   {\ensuremath{\pi^{-}}}
\newcommand{\piplus}  {\ensuremath{\pi^{+}}}
\newcommand{\kmin}    {\ensuremath{\textrm{K}^-}}
\newcommand{\kplus}   {\ensuremath{\textrm{K}^+}}
\newcommand{\pbar}    {\ensuremath{\bar{\textrm{p}}}}
\newcommand{\ommin}   {\ensuremath{\Omega^-}}
\newcommand{\omminbf} {\ensuremath{\mathbf{\Omega^-}}}        % bold, for title
\newcommand{\omplus}  {\ensuremath{\bar{\Omega}^+}}           % with bar
\newcommand{\omplusbf}{\ensuremath{\mathbf{\bar{\Omega}^+}}}  % bold with bar
\newcommand{\lamavg}  {\ensuremath{\langle \Lambda \rangle}}
\newcommand{\labavg}  {\ensuremath{\langle \bar{\Lambda} \rangle}}
\newcommand{\ximavg}  {\ensuremath{\langle \Xi^{-} \rangle}}
\newcommand{\xipavg}  {\ensuremath{\langle \bar{\Xi}^{+} \rangle}}
\newcommand{\omavg}   {\ensuremath{\langle \Omega \rangle}}
\newcommand{\piavg}   {\ensuremath{\langle \pi \rangle}}
\newcommand{\pipavg}  {\ensuremath{\langle \pi^{+} \rangle}}
\newcommand{\pimavg}  {\ensuremath{\langle \pi^{-} \rangle}}
\newcommand{\kpavg}   {\ensuremath{\langle \textrm{K}^+ \rangle}}
\newcommand{\jpsi}    {\ensuremath{\textrm{J}/\psi}}
\newcommand{\psip}    {\ensuremath{\psi^{\prime}}}
\newcommand{\chic}    {\ensuremath{\chi_{c}}}
\newcommand{\psitwos} {\ensuremath{\psi(\textrm{2S})}}
\newcommand{\lamp}    {\ensuremath{\textrm{p}-\Lambda}}
\newcommand{\lamlam}  {\ensuremath{\Lambda-\Lambda }}
\newcommand{\hzero}   {\ensuremath{\textrm{H}^0}}
\newcommand{\mpmm}    {\ensuremath{\mu^{+} \mu^{-}}}
\newcommand{\epem}    {\ensuremath{\textrm{e}^{+} \textrm{e}^{-}}}
\newcommand{\epm}     {\ensuremath{\textrm{e}^{\pm}}}
\newcommand{\elp}     {\ensuremath{\textrm{e}^{+}}}
\newcommand{\elm}     {\ensuremath{\textrm{e}^{-}}}
\newcommand{\qqbar}   {\ensuremath{\textrm{q}\bar{\textrm{q}}}}
\newcommand{\ccbar}   {\ensuremath{\textrm{c}\bar{\textrm{c}}}}
\newcommand{\ppbar}   {\ensuremath{\textrm{p}\bar{\textrm{p}}}}
\newcommand{\hyptri}  {\ensuremath{{}^{3}_{\Lambda}\textrm{H}}}
\newcommand{\hyptrib} {\ensuremath{{}^{3}_{\bar{\Lambda}}\overline{\textrm{H}}}}
\newcommand{\sqrts}   {\ensuremath{\sqrt{s}}}
\newcommand{\sqrtsnn} {\ensuremath{\sqrt{s_{_{\rbt{NN}}}}}}
\newcommand{\pT}      {\ensuremath{p_{\rb{T}}}}
\newcommand{\pt}      {\ensuremath{p_{\rb{t}}}}
\newcommand{\zt}      {\ensuremath{z_{\rb{T}}}}
\newcommand{\ptavg}   {\ensuremath{\langle p_{\rb{T}} \rangle}}
\newcommand{\ptavgsq} {\ensuremath{\langle p_{\rb{T}}^{2} \rangle}}
\newcommand{\mt}      {\ensuremath{m_{\rb{T}}}}
\newcommand{\kt}      {\ensuremath{k_{\rb{T}}}}
\newcommand{\mtmzero} {\ensuremath{m_{\rb{T}} - m_{\rb{0}}}}
\newcommand{\mzero}   {\ensuremath{m_{\rb{0}}}}
\newcommand{\mtavg}   {\ensuremath{\langle m_{\rb{T}} \rangle}}
\newcommand{\mtmavg}  {\ensuremath{\langle m_{\rb{T}} \rangle - m_{\rb{0}}}}
\newcommand{\minv}    {\ensuremath{m_{\rb{inv}}}}
\newcommand{\qinv}    {\ensuremath{q_{\rb{inv}}}}
\newcommand{\dedx}    {\ensuremath{\textrm{d}E/\textrm{d}x}}
\newcommand{\dndy}    {\ensuremath{\textrm{d}N/\textrm{d}y}}
\newcommand{\dndyjp}  {\ensuremath{\textrm{d}N_{\rb{J/}\psi}/\textrm{d}y}}
\newcommand{\dndymu}  {\ensuremath{\textrm{d}N_{\mu}/\textrm{d}y}}
\newcommand{\dndetach}{\ensuremath{\textrm{d}N_{\rb{ch}}/\textrm{d}\eta}}
\newcommand{\dndetatr}{\ensuremath{\textrm{d}N_{\rb{trk}}/\textrm{d}\eta}}
\newcommand{\dndpt}   {\ensuremath{\textrm{d}N/\textrm{d}p_{\rb{T}}}}
\newcommand{\nch}     {\ensuremath{N_{\rb{ch}}}}
\newcommand{\nhits}   {\ensuremath{N_{\rb{hits}}}}
\newcommand{\nhitsT}  {\ensuremath{N_{\rb{hits}}^{\rbt{TOF}}}}
\newcommand{\nhitsR}  {\ensuremath{N_{\rb{hits}}^{\rbt{RPC}}}}
\newcommand{\nhitsS}  {\ensuremath{N_{\rb{hits}}^{\rbt{TOF+RPC}}}}
\newcommand{\ntracks} {\ensuremath{N_{\rb{tracks}}}}
\newcommand{\nwound}  {\ensuremath{N_{\rb{w}}}}
\newcommand{\npart}   {\ensuremath{N_{\rb{part}}}}
\newcommand{\nspec}   {\ensuremath{N_{\rb{spec}}}}
\newcommand{\ncoll}   {\ensuremath{N_{\rb{coll}}}}
\newcommand{\taa}     {\ensuremath{T_{\rbt{AA}}}}
\newcommand{\npartav} {\ensuremath{\langle N_{\rb{part}} \rangle}}
\newcommand{\nspecav} {\ensuremath{\langle N_{\rb{spec}} \rangle}}
\newcommand{\ncollav} {\ensuremath{\langle N_{\rb{coll}} \rangle}}
\newcommand{\taaav}   {\ensuremath{\langle T_{\rbt{AA}} \rangle}}
\newcommand{\bav}     {\ensuremath{\langle b \rangle}}
\newcommand{\navg}    {\ensuremath{\langle N \rangle}}
\newcommand{\sinel}   {\ensuremath{\sigma_{\rb{inel}}}}
\newcommand{\tf}      {\ensuremath{T_{\rb{f}}}}
\newcommand{\tch}     {\ensuremath{T_{\rb{ch}}}}
\newcommand{\mub}     {\ensuremath{\mu_{\rbt{B}}}}
\newcommand{\gams}    {\ensuremath{\gamma_{\rb{s}}}}
\newcommand{\betat}   {\ensuremath{\beta_{\rb{T}}}}
\newcommand{\etaf}    {\ensuremath{\eta_{\rb{f}}}}
\newcommand{\tauzero} {\ensuremath{\tau_{\rb{0}}}}
\newcommand{\btavg}   {\ensuremath{\langle \beta_{\rb{T}} \rangle}}
\newcommand{\betas}   {\ensuremath{\beta_{\rb{s}}}}
\newcommand{\der}     {\ensuremath{\textrm{d}}}
\newcommand{\rsnn}    {\ensuremath{\sqrt{s_{\rb{NN}}}}}
\newcommand{\tpm}     {\ensuremath{\! \pm \!}}
\newcommand{\ebeam}   {\ensuremath{E_{\rb{beam}}}}
\newcommand{\yproj}   {\ensuremath{y_{\rb{proj}}}}
\newcommand{\ybeam}   {\ensuremath{y_{\rb{beam}}}}
\newcommand{\dzeros}  {\ensuremath{d_{\rb{0}}^{\rb{S}}}}
\newcommand{\dzerot}  {\ensuremath{d_{\rb{0}}^{\rb{T}}}}
\newcommand{\fzeros}  {\ensuremath{f_{\rb{0}}^{\rb{S}}}}
\newcommand{\fzerot}  {\ensuremath{f_{\rb{0}}^{\rb{T}}}}
\newcommand{\chisq}   {\ensuremath{\chi^{2}}}
\newcommand{\vtxz}    {\ensuremath{z_{\rb{vtx}}}}
\newcommand{\raa}     {\ensuremath{R_{\rbt{AA}}}}
\newcommand{\ptsqraa} {\ensuremath{r_{\rbt{AA}}}}
\newcommand{\lumint}  {\ensuremath{{\cal L}_{\rb{int}}}}
\newcommand{\acceff}  {\ensuremath{{A \times \epsilon}}}
\newcommand{\fb}      {\ensuremath{f_{\rbt{B}}}}
\newcommand{\fbprime} {\ensuremath{f_{\rbt{B}}^{\prime}}}
\newcommand{\fsig}    {\ensuremath{f_{\rbt{Sig}}}}
\newcommand{\ffb}     {\ensuremath{F_{\rbt{Bkg}}}}
\newcommand{\ffs}     {\ensuremath{F_{\rbt{Sig}}}}
\newcommand{\rzero}   {\ensuremath{R_{\rb{0}}}}
\newcommand{\rgaus}   {\ensuremath{R_{\rb{G}}}}
\newcommand{\rgeo}    {\ensuremath{R_{\rb{geo}}}}
\newcommand{\rout}    {\ensuremath{R_{\rb{out}}}}
\newcommand{\rside}   {\ensuremath{R_{\rb{side}}}}
\newcommand{\rlong}   {\ensuremath{R_{\rb{long}}}}
%
%===============================================================================

\title{Open Questions in the Understanding of Strangeness Production
  in HIC -- Experiment Perspective}
%
% subtitle is optional
\author{\firstname{Christoph}
  \lastname{Blume}\inst{1}\fnsep\thanks{\email{blume@ikf.uni-frankfurt.de}}}
\institute{Institut f\"{u}r Kernphysik, Goethe-Universit\"{a}t
  Frankfurt am Main,\\
  Max-von-Laue-Str.~1, 60438~Frankfurt am Main, Germany}
\abstract{Open questions concerning strangeness production in
  heavy-ion collisions are discussed with a focus on the experimental
  aspects.  The open points are presented in the context of recent
  measurements.}
\maketitle
\section{Introduction}
\label{sect:intro}

The following is an attempt to compile a list of the most important
open topics concerning strangeness production in heavy-ion physics.
This list is seen from an experimentalist point-of-view (the theory
perspective is discussed in \cite{chemingsqm17}) and is also naturally
incomplete and biased.  In order to not get lost in the many facets of
strangeness physics, the questions will be limited to those belonging
to four main subjects: energy dependence of strangeness enhancement in
nucleus-nucleus collisions, understanding of small systems,
strangeness production at low energies and
hyperon-interaction \& hypernuclei.

\section{Energy dependence of strangeness enhancement}
\label{sect:aa}

\begin{figure}[th]
\centering
\includegraphics[width=0.7\textwidth,clip]{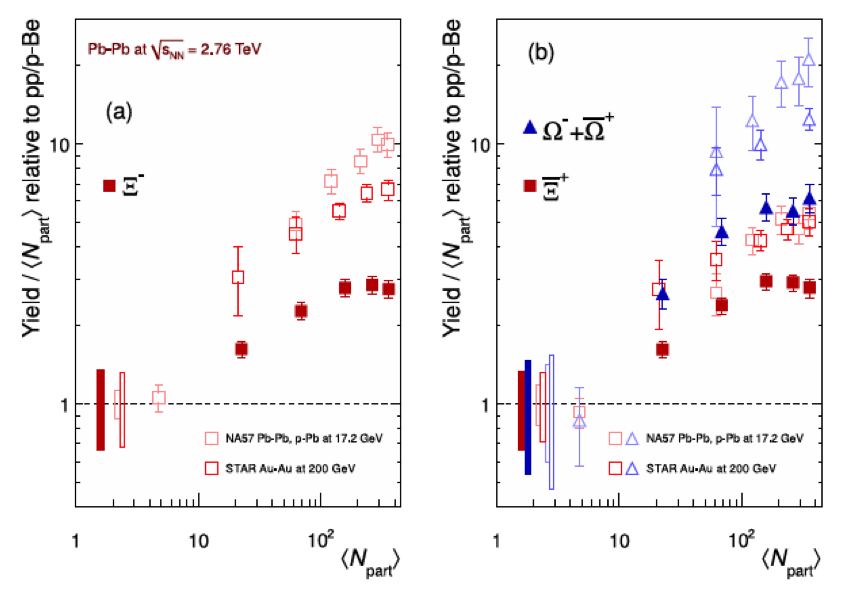}
\caption{The enhancement factors for multi-strange (anti-)particles as
  a function of the number of participants \npart\ as measured at the
  SPS \cite{Antinori:2006ij}, RHIC \cite{Abelev:2007xp} and LHC
  \cite{ABELEV:2013zaa}.}
\label{fig:strange_enh}
\end{figure}

\begin{figure}[th]
\centering
\includegraphics[width=0.43\textwidth,clip]{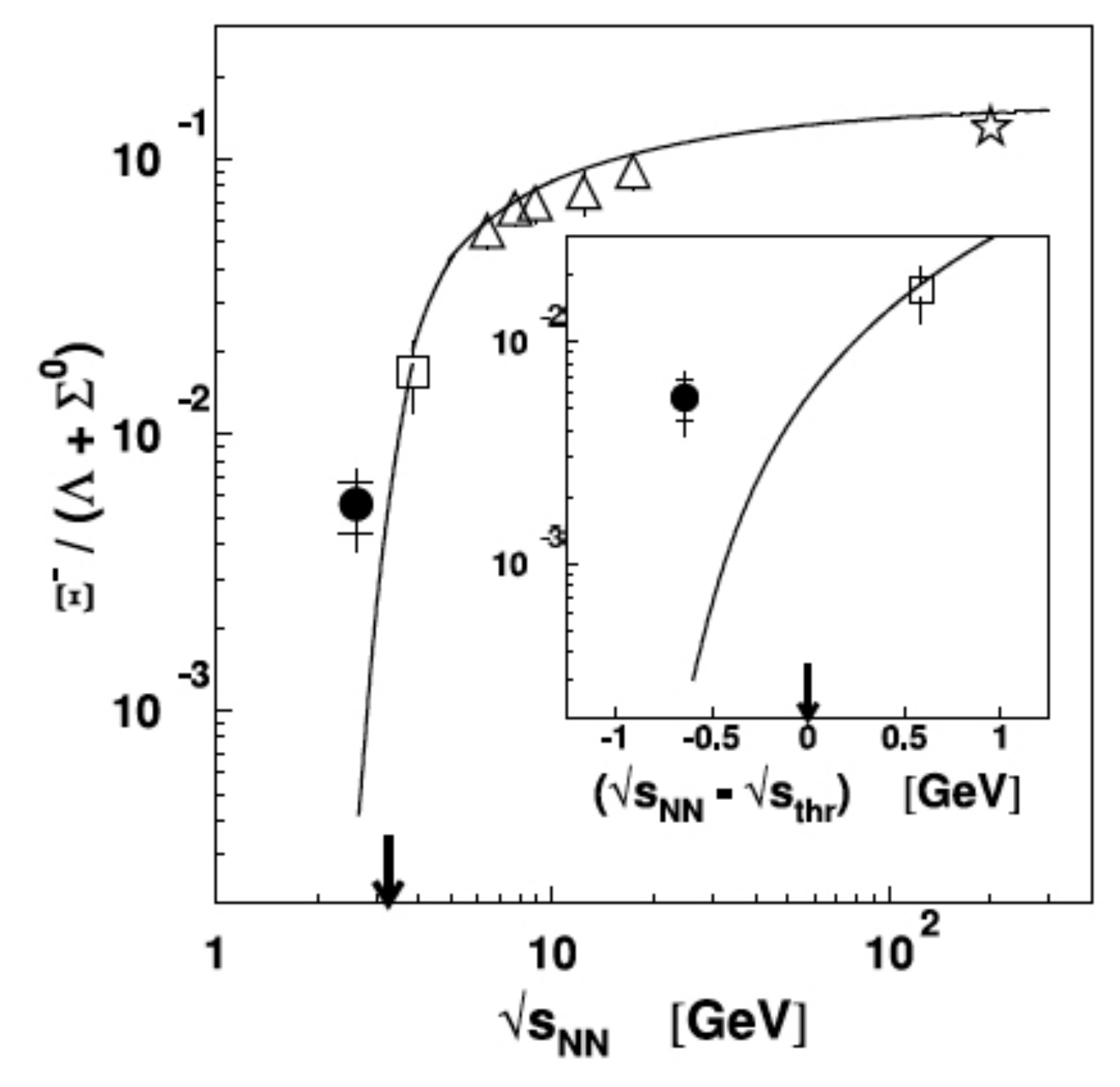}
\includegraphics[width=0.53\textwidth,clip]{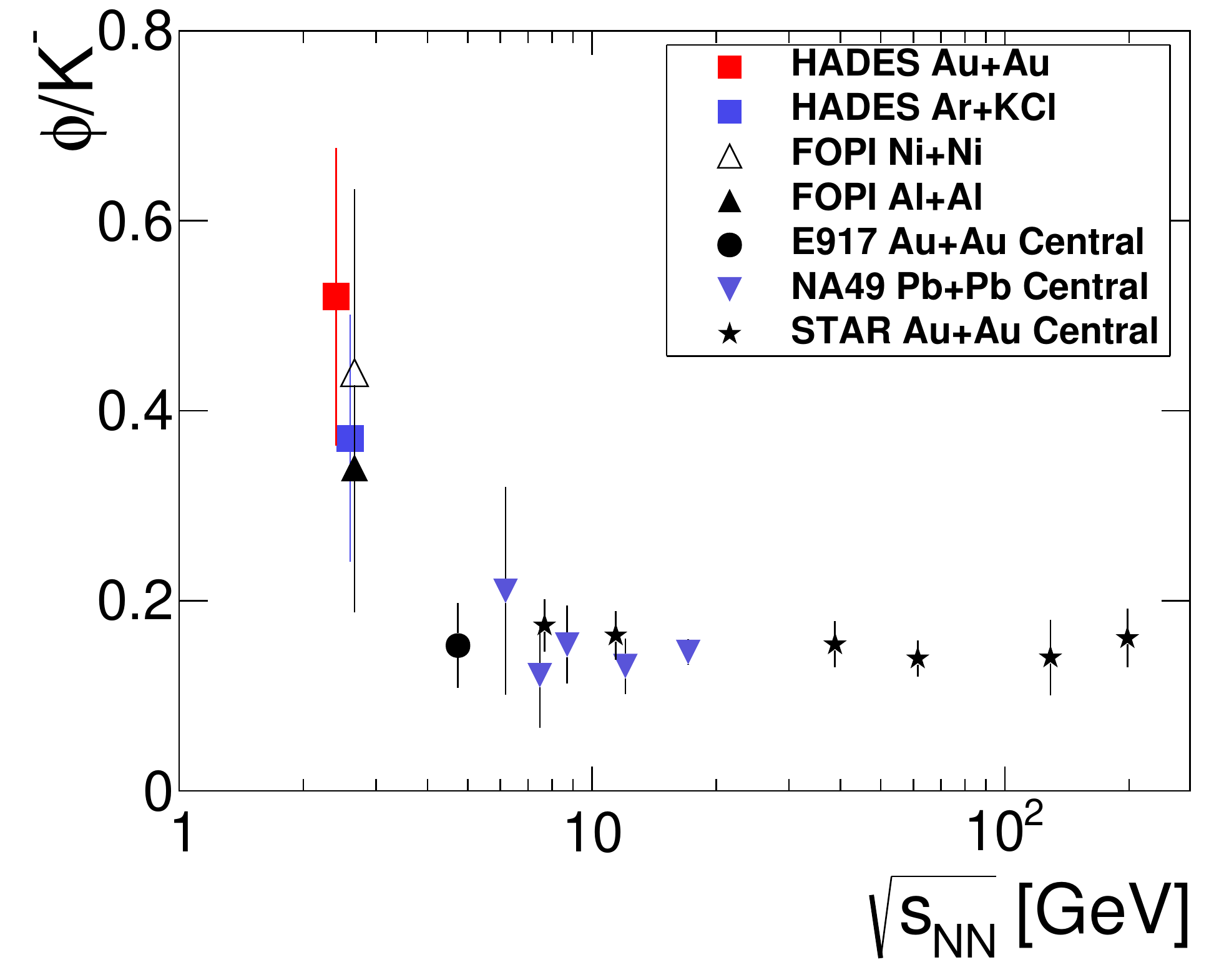}
\caption{The $\xim/(\lam + \sig)$-ratio (left panel,
  \cite{Agakishiev:2009rr}) and the $\phi/\kmin$-ratio (right panel,
  \cite{Adamczewski-Musch:2017rtf}) as a function of the
  centre-of-mass energy as measured in central heavy-ion collisions.}
\label{fig:xi_phi_lowe}
\end{figure}

It has been established already quite a while ago that the production
of strange particles is significantly enhanced in heavy-ion reactions
relative to elementary proton-proton collisions (for a review see
\cite{Blume:2011sb}).  Usually, the enhancement factor $E$ is defined
as:
\begin{equation}
E = \frac{2}{\npart}
       \left( \left. \frac{dN(\textrm{AA})}{dy} \right|_{y = 0} \right) \left/
       \left( \left. \frac{dN(\textrm{pp})}{dy} \right|_{y = 0}
       \right) \right.
\end{equation}
It is a remarkable fact that $E$ decreases significantly when going
from SPS energies to the very high energies available at the LHC.
E.g. for the \ommin\ $E$ is found to be around 20 at $\sqrtsnn =
17.3$~GeV \cite{Antinori:2006ij}, $\sim 12$ at $\sqrtsnn = 200$~GeV
\cite{Abelev:2007xp} and only $\sim 6$ at $\sqrtsnn = 2.76$~TeV
\cite{ABELEV:2013zaa}, as illustrated in Fig.~\ref{fig:strange_enh}.
While the enhancement factor has been measured for essentially all
multi-strange (anti-)particles at these energies, data are still quite
scarce for these rare particles at energies below $\sqrtsnn =
17.3$~GeV.  An interesting exception is the measurement of \xim\
production in Ar+KCl collisions at 1.76\agev\ beam energy by the HADES
collaboration (left panel of Fig.~\ref{fig:xi_phi_lowe})
\cite{Agakishiev:2009rr}.  HADES also has studied the production of
$\phi$ mesons at these sub-threshold energies (right panel of
Fig.~\ref{fig:xi_phi_lowe}) \cite{Adamczewski-Musch:2017rtf}.  The
data are in so far remarkable as also here an enhancement of the rare
strange particles is observed.  The $\xim/(\lam + \sig)$-ratio is
found to be much larger than the statistical model expectation and the
$\phi/\kmin$-ratio rises dramatically towards very low energies.

\begin{figure}[th]
\centering
\includegraphics[width=0.60\textwidth,clip]{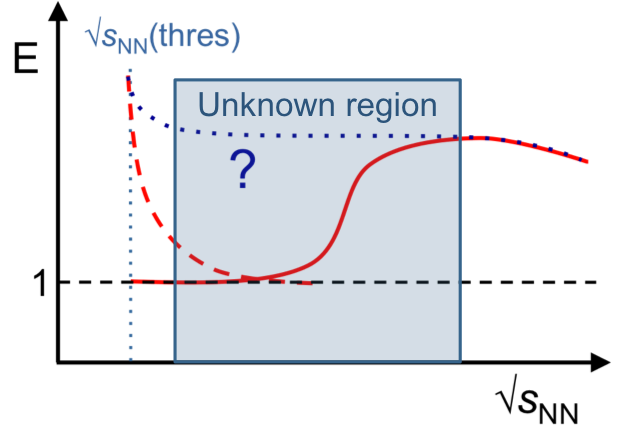}
\caption{Sketch of the energy dependence of the enhancement factor of
  rare multi-strange (anti-)particles.}
\label{fig:enh_edep}
\end{figure}

Taking these observations together, one finds that the strangeness
enhancement, in particular for rare multi-strange (anti-)particles,
exhibits a quite complicated energy dependence.  As sketched in
Fig.~\ref{fig:enh_edep} there is evidence for a strong increase at
sub-threshold energies and it has been established that there is a
strong enhancement at intermediate (i.e. SPS energies) energies, which
slowly decreases again towards high energies (i.e. RICH and LHC).  It
remains an open question if there is any evidence for an onset of
strangeness enhancement between low and intermediate energies.  This
should happen in a region where sub-threshold phenomena (dashed line)
do not play a role any more and the partonic degrees of freedom (solid
line) are becoming more and more effective.  It might also very well
be that there is no discernible onset visible, due to additional
hadronic medium effects (dotted line) (e.g. multi-step hadronic
reactions, resonances, multi-meson fusion processes, etc.).  However,
it is obvious that high quality data of rare particles in the energy
range below $\sqrtsnn \sim 10$~GeV would be highly relevant (see also
\cite{dansqm17}) in order to answer the following questions:

\begin{itemize}

\item What is the energy dependence of strangeness enhancement over
  the whole energy region, in particular for multi-strange
  (anti-)particles?

\item Could there be an onset somewhere?

\item To what extend can hadronic effects cause a strangeness
  enhancement at intermediate energies (SPS and below)?

\item Do we understand the dramatic effects at sub-threshold energies?

\item Or, in other words, can finally a direct connection between
  strangeness enhancement and QGP formation be established?

\end{itemize}

\section{Understanding of small systems}
\label{sect:small_system}

\begin{figure}[th]
\centering
\includegraphics[width=0.43\textwidth,clip]{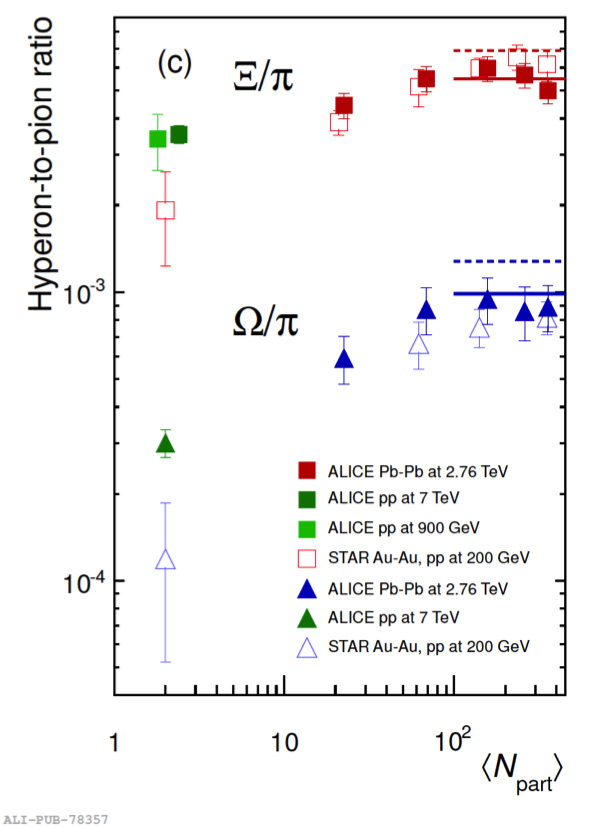}
\includegraphics[width=0.37\textwidth,clip]{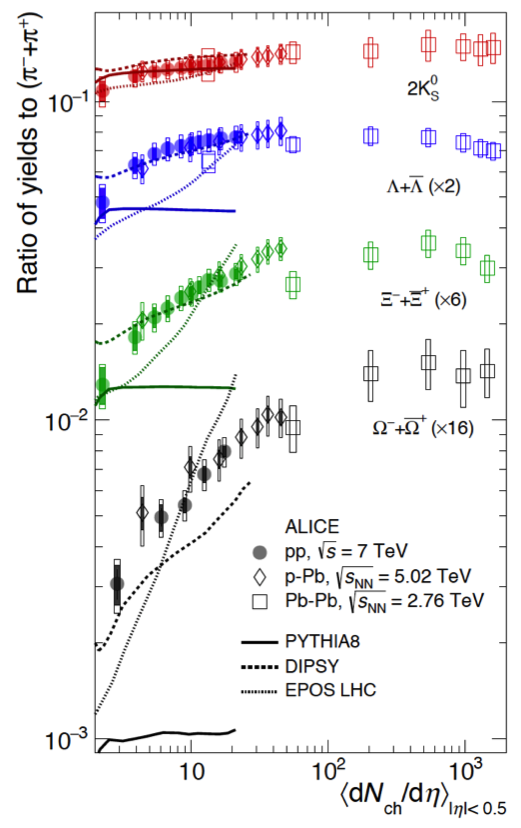}
\caption{Left: the $\xim/\pi$ and $\Omega/\pi$ ratios in pp and
  AA collisions as a function of \npartav\ \cite{ABELEV:2013zaa}.  The
  lines correspond to statistical model predictions (solid line
  \cite{Andronic:2008gu}, dashed line \cite{Cleymans:2006xj}).
  Right: the yield ratios of strange particles and pions measured in
  pp, pA and AA collisions at the LHC as a function of \dndetach\
  \cite{ALICE:2017jyt}.}
\label{fig:systemsize_enh}
\end{figure}

\begin{figure}[th]
\centering
\includegraphics[width=0.32\textwidth,clip]{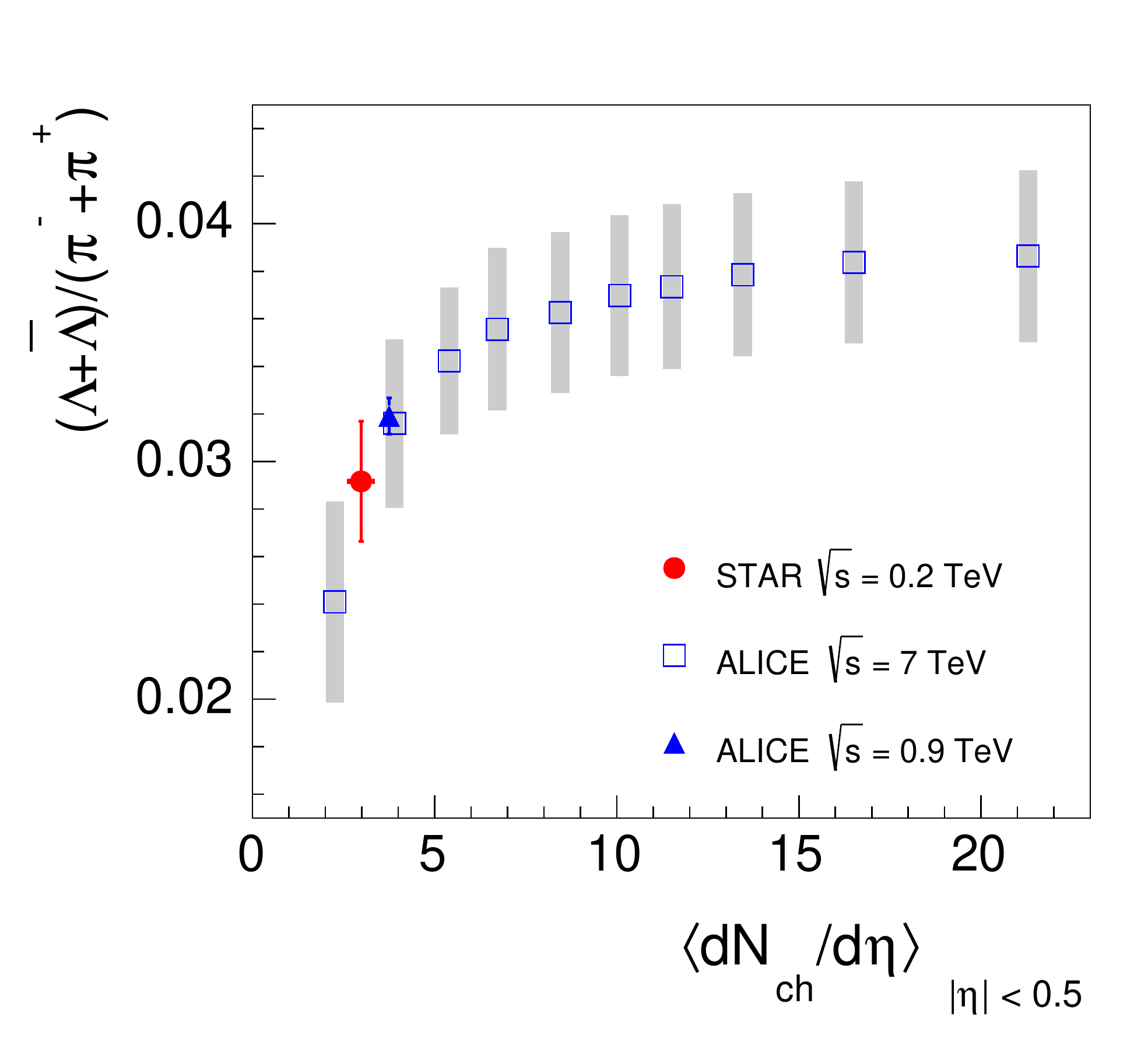}
\includegraphics[width=0.32\textwidth,clip]{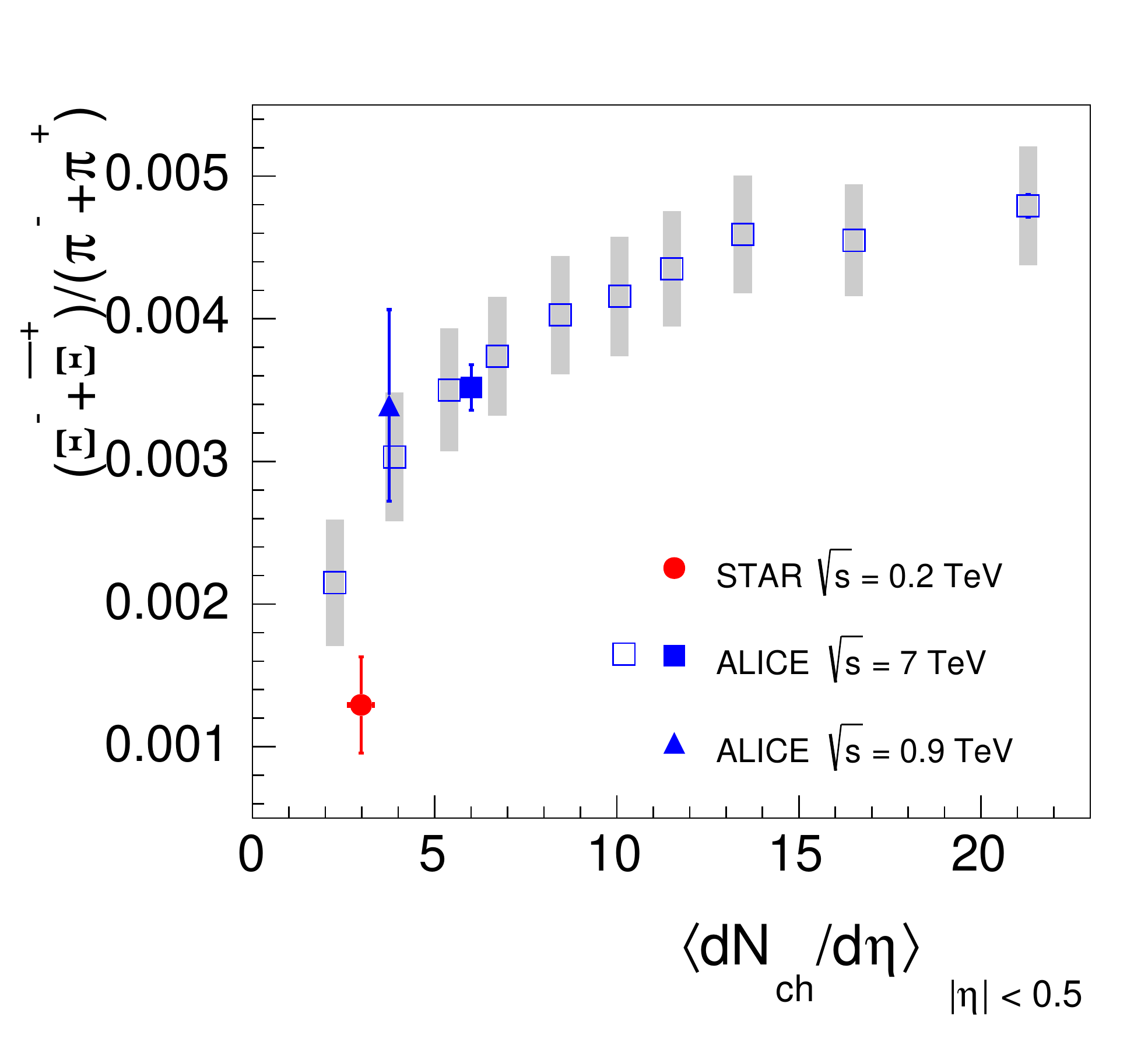}
\includegraphics[width=0.32\textwidth,clip]{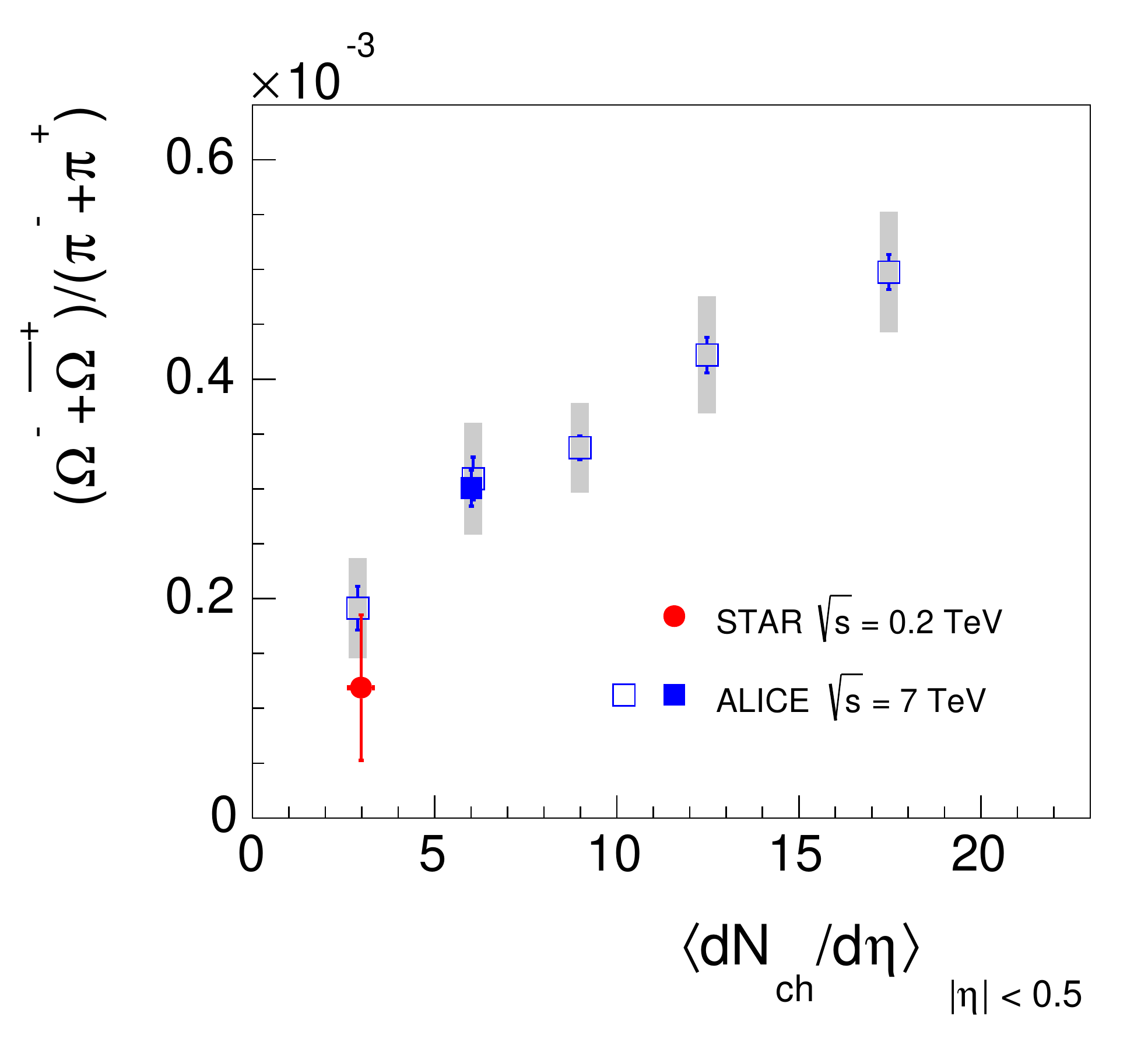}
\caption{The yield ratios of strange particles to pions measured in
  multiplicity selected pp collisions at $\sqrts = 7$~TeV
  \cite{ALICE:2017jyt} and in minimum bias pp collisions measured at
  various energies by ALICE \cite{Abelev:2012jp,Aamodt:2011zza} and
  STAR \cite{Abelev:2006cs,Abelev:2008ab} as a function of \dndetach.}
\label{fig:mult_dep_lmxiom}
\end{figure}

Small systems play a key role for the understanding of strangeness
production in heavy-ion collisions.  The decrease of the strangeness
enhancement factor between SPS and LHC energies, as discussed in the
previous section, is actually not caused by a decrease of the yields
in heavy-ion collisions.  In fact, they are already close to the
statistical model expectation at top SPS-energies and increase only
slightly towards higher energies.  However, the yields of
multi-strange particles increases much stronger in pp collisions than
in AA (see left panel of Fig.~\ref{fig:systemsize_enh}).  For
instance, one finds that the $\Xi/\pi$ ratios are almost the same at
RHIC and LHC in central AA collisions, while this ratios increase
significantly in pp collisions.  Thus, the decrease of the strangeness
enhancement is due to a release of the strangeness suppression in pp with
increasing energies.

This phenomena can also be investigated by comparing pp collisions of
different reaction violence at high energies.  This is done by
selecting pp event classes of different charged particle multiplicity
\dndetach.  Such a study has recently be performed by the ALICE
collaboration \cite{ALICE:2017jyt} (see right panel of
Fig.~\ref{fig:mult_dep_lmxiom}).  Here it was found that the yield
ratios of strange particles to pions observed in very high multiplicity
pp collisions are on the same level as the ones measured in peripheral
heavy-ion reactions.  Also, the multiplicity dependence of these
ratios turns out to be very similar in pp and pA collisions. This is
not at all trivial since the physics in a high multiplicity pp
collision is quite different from the one in pA at the same
multiplicity.  While in the first case a very rare and violent pp
interaction has to be involved, in the latter case the same
multiplicity can be achieved by the much more likely superposition of
several soft pp collisions.

In the statistical model approach strangeness enhancement is
described by the transition from a canonical to a grand-canonical
ensemble, which depends on the volume $V_{0}$ of the system.  In this
picture a strangeness hierarchy is expected \cite{Hamieh:2000tk},
i.e. the volume dependence gets stronger with increasing strangeness
content.  Since such a hierarchy was observed in the multiplicity
dependence \cite{ALICE:2017jyt}, it is natural to interpret the pp
data in this way by assuming a relation $V_{0} \propto
\langle\dndetach\rangle$.  In fact, as reasonable description of all
particle ratios, with the notable exception of the $\phi/\pi$-ratio
(other peculiarities related to the $\phi$-meson are discussed in
Sect.~\ref{sect:low_energies}), can be achieved within this model
\cite{Vislavicius:2016rwi}.

If this interpretation holds, the particle ratios should saturate for
pp collisions at very high multiplicities when the grand-canonical
limit is reached.  As shown in Fig.~\ref{fig:mult_dep_lmxiom}, this
might indeed already be the case for the $\lam/\pi$ and $\Xi/\pi$
ratios, while for the $\Omega/\pi$ it is rather still a continuous
increase.  Also, it would be interesting to establish whether
\dndetach\ does provide an universal scaling variable for all
energies.  Fig.~\ref{fig:mult_dep_lmxiom} shows a comparison of the
multiplicity selected pp collisions at $\sqrtsnn = 2.76$~TeV to
minimum bias pp reactions at lower energies.  The minimum bias data
points roughly follow the trend, however, it would be worthwhile to
examine the multiplicity dependence also at low energies with good
accuracy, in order to test whether really an universal scaling holds.
In summary, the following questions would require further
investigations:

\begin{itemize}

\item Does \dndetach\ provide an universal scaling for system size
  dependencies (pp $\rightarrow$ pA $\rightarrow$ AA)?

\item Is the relation of \dndetach\ to the reaction volume the only
  relevant factor (look at other observables)? 

\item Does the multiplicity dependence match the transition from
  a canonical to a grand-canonical ensemble at all energies?

\item Is a saturation of particle ratios observed in very high
  multiplicity pp collisions?

\end{itemize}

\section{Low energies and the $\phi$-meson}
\label{sect:low_energies}

\begin{figure}[th]
\centering
\includegraphics[width=0.43\textwidth,clip]{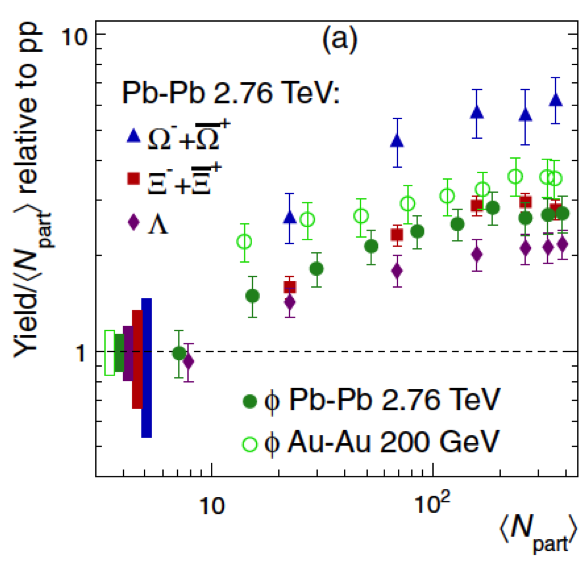}
\includegraphics[width=0.43\textwidth,clip]{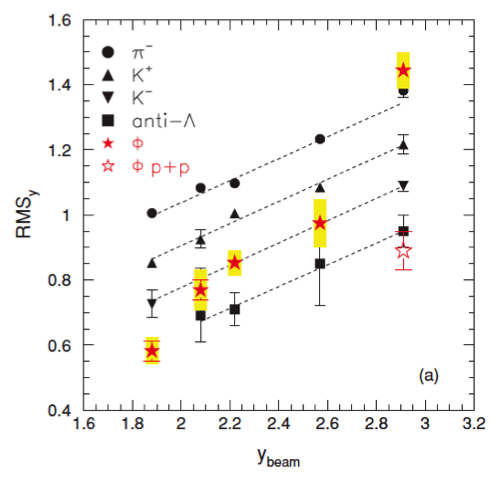}
\caption{Left: the yield ratios of strange particles to pions measured
  in pp and AA collisions at different centre-of-mass energies
  \cite{Abelev:2014uua}.
  Right: the widths of the rapidity distributions of the $\phi$-meson
  measured at the SPS as a function of the beam rapidity \ybeam\
  \cite{Alt:2008iv}.}
\label{fig:phi}
\end{figure}

Also at the lower end of the energy spectrum many unresolved
questions are remaining.  These are particularly important, since they
are concerned with the behaviour of strange particles in a hadronic
medium, which need to be understood before any conclusions about the
partonic phases of heavy-ion collisions can be made.

A topic discussed since quite a while is the propagation of kaons in
the medium.  In low energy AA collisions a significant difference
between the inverse slope parameters of the \pt~spectra of \kmin\ and
\kplus\ has been observed ($T(\kplus) > T(\kmin)$).  Since the cross
section of reactions with nucleons are different ($\sigma(\kmin) >
\sigma(\kplus)$) due to strangeness exchange reactions of the \kmin,
the rescattering with the hadronic medium should cause different
freeze-out conditions for the two kaon species and thus can explain
the different slope parameters \cite{Hartnack:2011cn}.  Also, the
kaon-nucleon potential, which is expected to be attractive for \kmin\
and repulsive for \kplus, can modify the inverse slope parameters.  On
the other side, the unexpectedly high $\phi/\kmin$-ratio at low
energies (see right panel of Fig.~\ref{fig:xi_phi_lowe}) can also
provide a natural explanation for this observation.  Since the
feed-down from $\phi$-decays at these energies is a substantial
contribution to the kaon spectra, it will also modify their shape.
This affects more strongly the rarer \kmin\ than \kplus\ and thus will
result in the different spectral shapes \cite{heidisqm17}.

With its $\textrm{s}\overline{\textrm{s}}$ valence quark structure the
$\phi$-meson is effectively a strangeness neutral particle ($S = 0$).
Nevertheless, it behaves in many ways as if it would have a non-zero
strangeness.  The enhancement factor for the $\phi$ is found to be
between the ones for the $\Lambda$ and $\Xi$ (see left panel of
Fig.~\ref{fig:phi}), thus rather behaving as a particle with an
effective strangeness in the range $1 - 2$.  Another so far
unexplained observation related to the $\phi$ is the energy dependence
of the widths of its rapidity distributions (right panel of
Fig.~\ref{fig:phi}).  In heavy-ion collisions, it broadens much
stronger than for $\pi$, \kmin\ and \kplus, which is difficult to
reconcile with kaon-coalescence being the main production mechanism
for $\phi$-mesons \cite{Alt:2008iv}.  Some important points to be
clarified concerning heavy-ion collisions at low energies and the
$\phi$-mesons are therefore:

\begin{itemize}

\item Do we understand the production and propagation of strangeness
  at low energies?

\item Is there any evidence for a sequential freeze-out due to
  different cross sections?

\item Does the medium also at low energy behave macroscopically and
  can fully be described by the statistical model?

\item Why does a non-strange particle behave so strange?

\end{itemize}

\section{Hyperon interaction and hypernuclei}
\label{sect:hyper}

\begin{figure}[th]
\centering
\includegraphics[width=0.44\textwidth,clip]{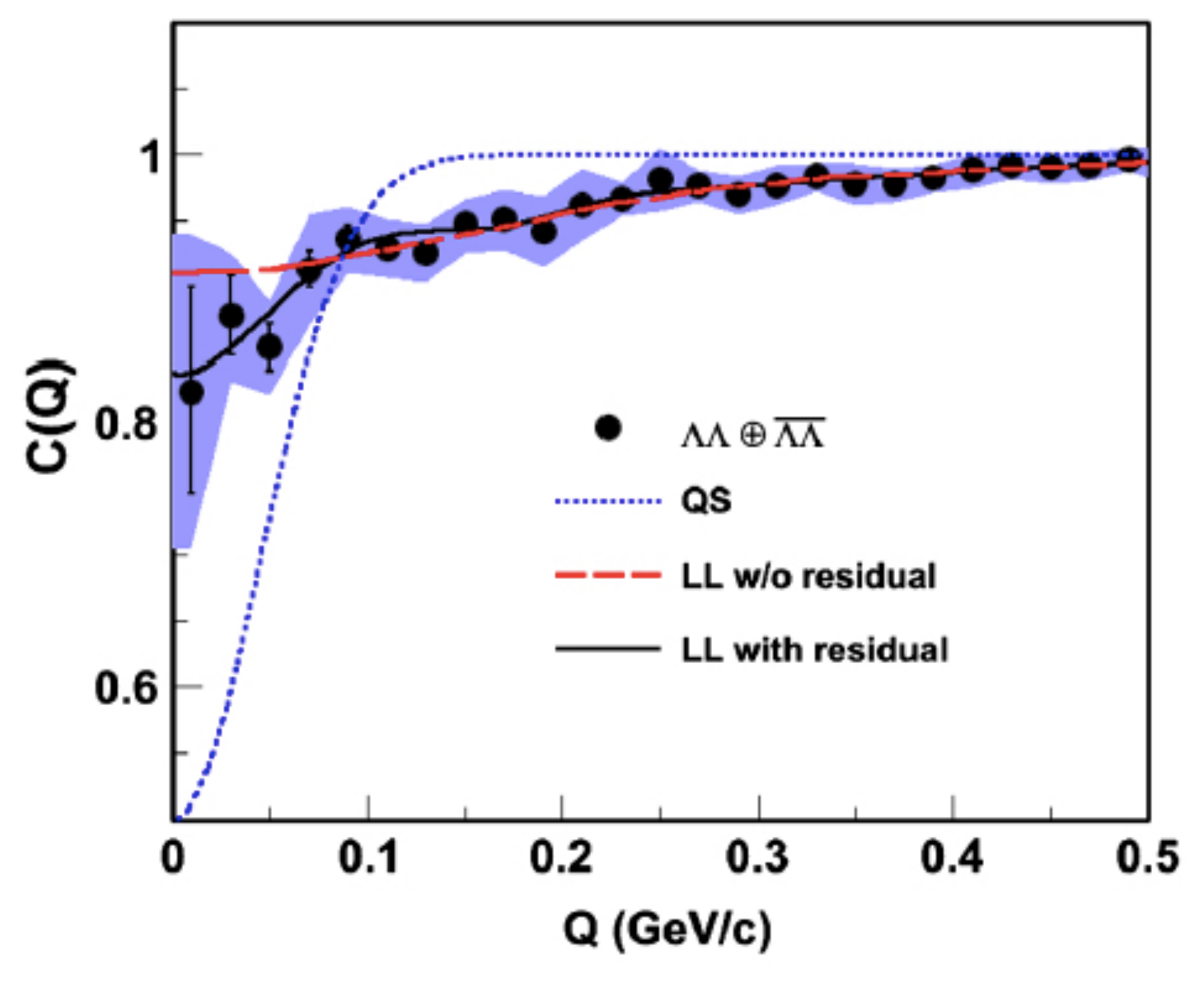}
\includegraphics[width=0.43\textwidth,clip]{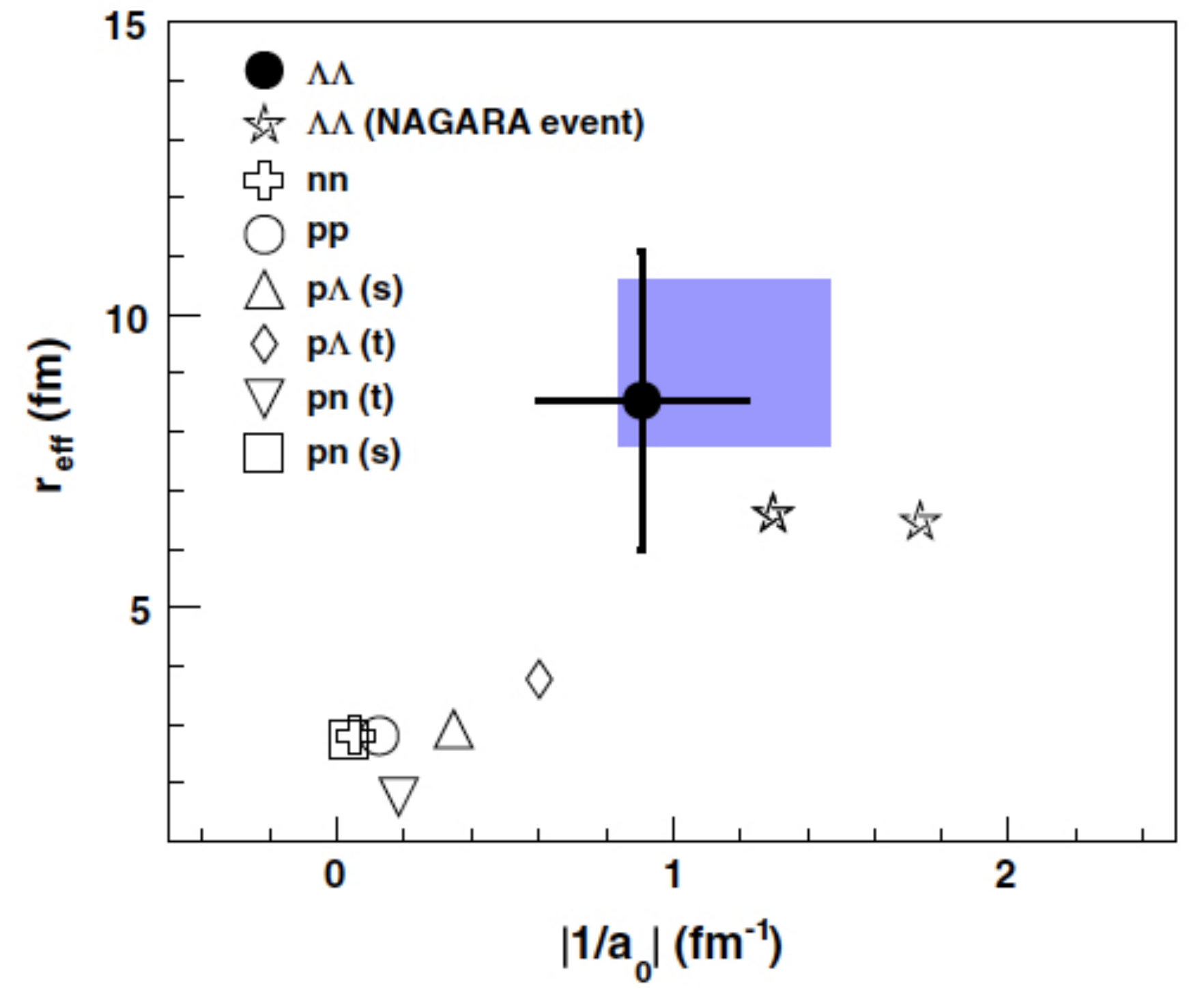}
\caption{Left: the $\lam\lam$-correlation function measured in Au+Au
  collisions at RHIC \cite{Adamczyk:2014vca}.  
  Right: the S-wave scattering length $a_{0}$ and effective
  interaction range $r_{\rb{eff}}$, as extracted from this correlation
  function with the model by Lednick\'{y} and Lyuboshitz
  \cite{Lednicky:1981su}.}
\label{fig:lamlam}
\end{figure}

\begin{figure}[th]
\centering
\includegraphics[width=0.43\textwidth,clip]{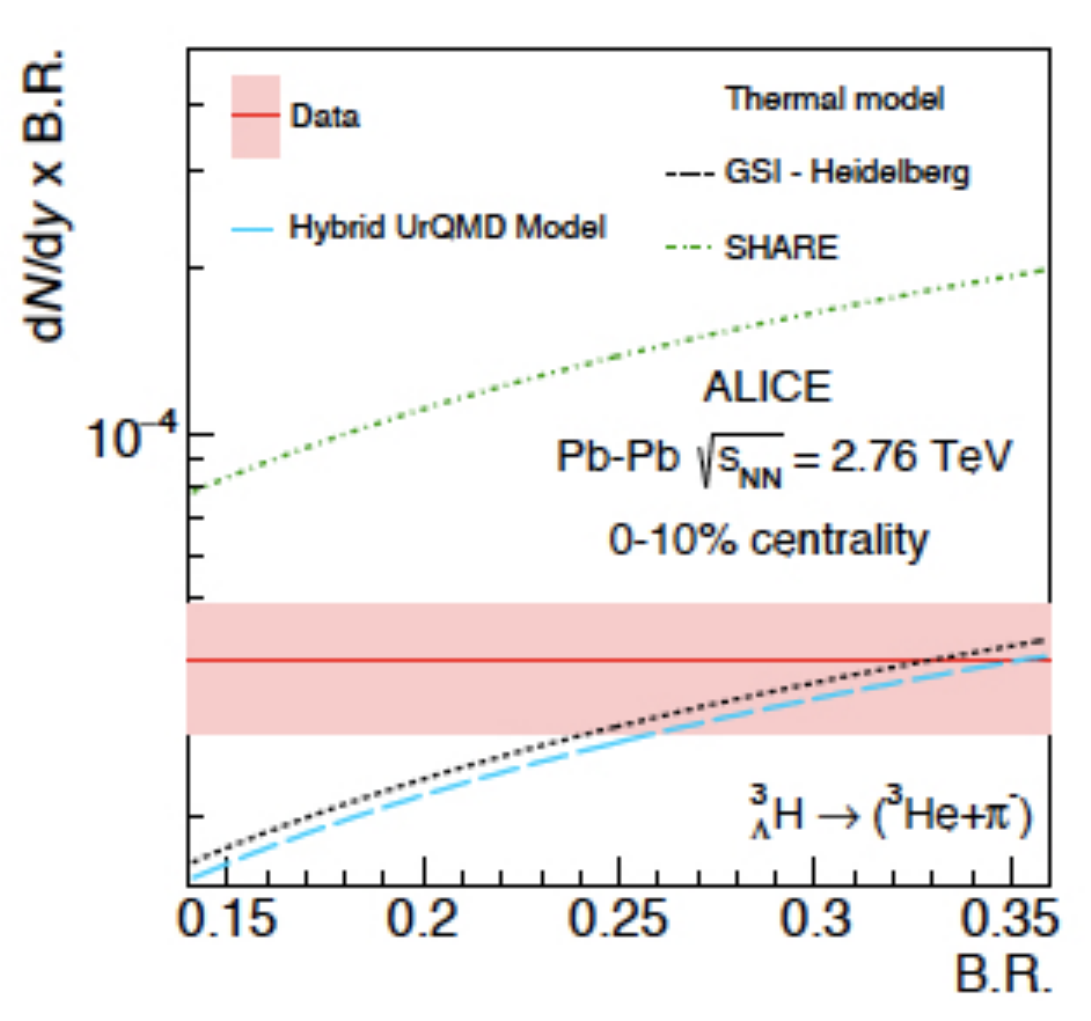}
\includegraphics[width=0.45\textwidth,clip]{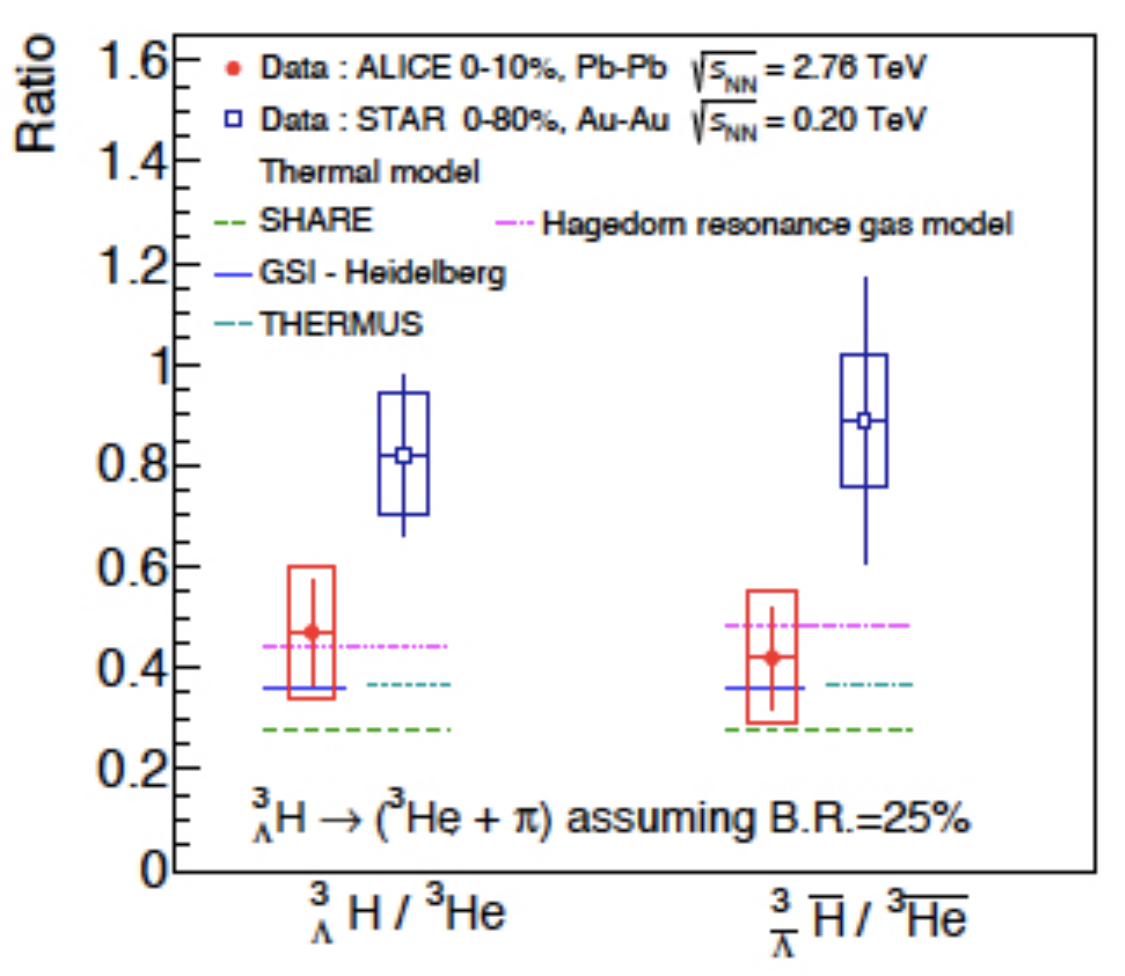}
\caption{Left: the \hyptri\ yield as measured at the LHC in
  comparison to several model predictions \cite{Adam:2015yta}.
  Right: the \hyptri/$^{3}$He and
  \hyptrib/$^{3}\overline{\textrm{He}}$ ratios measured at RICH and
  LHC in comparison to statistical model predictions
  \cite{Adam:2015yta}.}
\label{fig:hyper}
\end{figure}

The investigation of hyperon-interactions is a crucial ingredient for
the theoretical description of neutron stars.  It may in particular be
relevant for the understanding of high mass neutron stars with $M >
2\;M_{\odot}$.  One way to obtain informations is via two-particle
correlations.  The STAR collaboration recently managed to extract a
$\Lambda\Lambda$-correlation function in heavy-ion collisions
\cite{Adamczyk:2014vca} (see left panel of Fig.~\ref{fig:hyper}).  As
the strong interaction between the $\Lambda$-pairs causes a deviation
of the correlation function from the quantum-statistical expectation
of $C_{\Lambda\Lambda}(Q = 0) = 0.5$, one can infer information on the
scattering lengths and effective interaction ranges by comparing it to
corresponding models.  Using the one by Lednick\'{y} and Lyuboshitz
\cite{Lednicky:1981su}, a weak repulsive interaction was inferred by
the STAR collaboration \cite{Adamczyk:2014vca}.  However, an
alternative analysis \cite{Morita:2014kza} rather favours a weak
attractive interaction.

Another important source of information on this subject are
hypernuclei and more and more data is becoming available.  For
instance, with the measurement of anti-hypertritons the STAR
collaboration achieved the first observation of an anti-hypernucleus
\cite{Abelev:2010rv}.  Generally, it is found that the
(anti-)hypertriton yields agree very well with statistical model
expectations \cite{Adam:2015yta}.  Since their binding energy is very
small, it is surprising that their yields are fixed in a chemical
freeze-out environment with a temperature higher by about two orders
of magnitude.  Current measurements of the $\Lambda$-lifetime from the
decay of \hyptri\ seem to indicate that it is slightly lower than the
one of a free $\Lambda$, which would be an indication for a
modification of hyperon properties inside a nuclear medium.

The study of hyperon-interactions and their properties in nuclei has
regained quite some momentum recently with the measurements at RICH
and LHC.  High statistics data, in particular on double-hypernuclei,
which will become available in the near future with facilities such as
FAIR, will allow to address the questions listed below with much more
precise information:

\begin{itemize}

\item What do we really know about hyperon-hyperon interactions?

\item What is the possible contribution to the understanding of
  large-mass neutron stars?

\item Why are the yields of very weakly bound objects (e.g. \hyptri)
  so well described by the statistical model (``snowball in hell'')?

\item Are the properties of hyperons modified inside nuclei?

\end{itemize}

\section{Acknowledgements}

The author would like to acknowledge many helpful discussions with
B.~D\"{o}nigus, D.~Cebra, C.-M.~Ko, M.~Lorenz, H.~Str\"{o}bele and
many others.

%
% BibTeX or Biber users please use (the style is already called in the
% class, ensure that the "woc.bst" style is in your local directory)
%
\bibliography{utrecht2017_blume}

\end{document}